\documentclass[aps,preprint,preprintnumbers,nofootinbib,
groupedaddress,superscriptaddress,amsmath,amssymb]{revtex4}
\usepackage{graphicx}
\usepackage{dcolumn}
\usepackage{bm}
\usepackage{amssymb}
\usepackage{amsmath}
\usepackage{epsfig}    
\usepackage{color}
\usepackage{slashed}
\usepackage{hhline}

\def\be{\begin{equation}}
\def\ee{\end{equation}}
\newcommand{\bea}{\begin{eqnarray}}
\newcommand{\eea}{\end{eqnarray}}

\begin{document}

{\begin{flushright}{KIAS-P16051}
\end{flushright}}

\title{ Explanation of $B\to K^{(*)} \ell^+ \ell^-$  and muon $g-2$, \\
and implications at the LHC}
%
%
\author{Chuan-Hung Chen}
\email{physchen@mail.ncku.edu.tw}
\affiliation{Department of Physics, National Cheng-Kung University, Tainan 70101, Taiwan}

\author{Takaaki Nomura}
\email{nomura@kias.re.kr}
\affiliation{School of Physics, KIAS, Seoul 130-722, Korea}

\author{Hiroshi Okada}
\email{macokada3hiroshi@cts.nthu.edu.tw}
\affiliation{Physics Division, National Center for Theoretical Sciences, Hsinchu 300, Taiwan}

\date{\today}

\begin{abstract}

 More than $3\sigma$ deviations from the standard model are observed in the angular observable $P'_5$ of $B\to K^* \mu^+ \mu^-$ and muon $g-2$. To resolve these anomalies, we extend the standard model by adding  two leptoquarks. It is found that the signal strength of the diphoton Higgs decay can exhibit a significant deviation from unity and is within the data errors. Although  $\ell_i \to \ell_j \gamma$ puts severe bounds on some couplings,  it is found that the excesses of $P'_5$ and muon $g-2$ can still be explained and can be accommodated to the measurement of  $B_s\to \mu^+ \mu^-$ in this model.  In addition,  the leptoquark effects can also explain the LHCb measurement of $R_K = BR(B^+ \to K^+ \mu^+\mu^-)/BR(B^+ \to K^+ e^+e^-)=0.745^{+0.090}_{-0.074} \pm 0.036$, which shows a $2.6\sigma$ deviation from the standard model prediction. 
 
\end{abstract}
\maketitle

\section{Introduction}

The standard model (SM) has been tested at an unprecedented  level of precision through various experiments. However, some excesses have not yet been completely resolved. The first case is the muon anomalous magnetic moment {(muon $g-2$)}, where the discrepancy  between experimental data and the SM prediction is currently $\Delta a_\mu = a^{\rm exp}_\mu - a^{\rm SM}_\mu =(28.8\pm 8.0)\times 10^{-10}$~\cite{PDG}. The second case is  the angular observable $P'_5$ of $B\to K^* \mu^+ \mu^-$~\cite{DescotesGenon:2012zf}, where  a $3.4\sigma$ deviation, resulting from  the integrated luminosity of 3.0 fb$^{-1}$ at the LHCb~\cite{Aaij:2015oid}, recently  confirmed an earlier result with  $3.7\sigma$ deviations~\cite{Aaij:2013qta}. Moreover, the same measurement with $2.1\sigma$ deviations was reported by Belle~\cite{Abdesselam:2016llu}.  
Also, the other relevant $P_i$ observables are defined in Ref.~\cite{Matias:2012xw}.
Various possible resolutions to this excess have been widely studied~\cite{Descotes-Genon:2013wba,Gauld:2013qja,Datta:2013kja,Hurth:2013ssa,Descotes-Genon:2014uoa,Altmannshofer:2014rta,Crivellin:2015mga,Sahoo:2015wya,Straub:2015ica,Crivellin:2015era,Lee:2015qra,Alonso:2015sja,Sahoo:2015qha,Sahoo:2015pzk,Chiang:2016qov,Belanger:2015nma,Dorsner:2016wpm,Becirevic:2015asa,Descotes-Genon:2015uva,Boucenna:2016wpr,Hiller:2016kry}.  The third case is the ratio  $R_K = BR(B^+\to K^+\mu^+\mu^-)/BR(B^+\to K^+ e^+e^-)$, where $BR(B^+ \to K^+ \ell^+ \ell^-)$ is the branching ratio (BR) of the decay $B^+ \to K^+ \ell^+ \ell^-$; and the LHCb measurement shows a $2.6\sigma$ deviation from the SM result~\cite{Aaij:2014ora}.  In order to explain the deviation, various mechanisms have been  proposed~\cite{Crivellin:2015mga,Hiller:2014yaa,Hurth:2014vma,Gripaios:2014tna,Glashow:2014iga,Sahoo:2015fla,Bauer:2015knc,Das:2016vkr,Li:2016vvp,Becirevic:2016yqi,Sahoo:2016pet,Bhattacharya:2016mcc,Duraisamy:2016gsd}.

 In addition to the  excesses mentioned above, the  LHC with energetic $pp$ collisions  can also be a good place to test the SM and provide possible excess signals.  For instance,  a hint of  resonance with a mass of around 750 GeV in the diphoton invariant mass spectrum was indicated by the ATLAS~\cite{Aaboud:2016tru} and CMS~\cite{Khachatryan:2016hje} experiments. Due to the results,   various proposals have been broadly proposed and studied~\cite{Harigaya:2015ezk,Backovic:2015fnp,Angelescu:2015uiz,Nakai:2015ptz,Buttazzo:2015txu,DiChiara:2015vdm,Knapen:2015dap,Pilaftsis:2015ycr,Franceschini:2015kwy,Ellis:2015oso,Gupta:2015zzs,Kobakhidze:2015ldh,Falkowski:2015swt,Benbrik:2015fyz,Wang:2015kuj,Dev:2015isx,Allanach:2015ixl,Wang:2015omi,Chiang:2015tqz,Huang:2015svl,Ko:2016wce,Nomura:2016seu,Kanemura:2015bli,Cheung:2015cug,Nomura:2016fzs,Ko:2016lai,Belanger:2016ywb,Dorsner:2016ypw,Han:2016bvl,Mambrini:2015wyu}. Although it turns out  that the resonance has not been confirmed by the updating data of ATLAS~\cite{ATLAS:2016eeo} and CMS~\cite{CMS:2016crm} and has been shown to be more like a statistical fluctuation, the search for the new exotic events in the LHC still continue and is an  essential mission. 

To resolve the excesses in a specific framework, we propose the extension of the SM by including leptoquarks (LQs), where the LQs are colored scalars that simultaneously couple to  the leptons and quarks.   Hence, the $b\to s \ell^+ \ell^-$ decays can arise from the tree-level LQ-mediated Feynman diagrams when the muon $g-2$ is induced from LQ loops.

In addition to the decays $B\to K^{(*)} \ell^+ \ell^-$, the effective interactions for $b\to s \ell^+ \ell^-$ can also contribute to $B_s \to \mu^+ \mu^-$, where the  BR, measured by  LHCb and CMS~\cite{CMS:2014xfa}, is given as:
\begin{equation}
\text{BR}(B_s \to \mu^+ \mu^-)^{\text exp}  = (2.8^{+0.7}_{-0.6}) \times 10^{-9} \quad \quad \text{(LHCb-CMS)}\,.
\end{equation}
We note that the dominant effective couplings for $b\to s \ell^+ \ell^-$ processes  are denoted by the Wilson coefficients $C_{9,10}$. Usually, both $C_9$ and $C_{10}$ are strongly correlated.  Since this experimental result is consistent with the SM prediction of $\text{BR}(B_s \to \mu^+ \mu^-)^{\text SM} \approx 3.65\times 10^{-9}$~\cite{Bobeth:2013uxa}, in order to accommodate the anomalies of $P'_5$ and $R_K$ to the measurement of  $B_s \to \mu^+ \mu^-$, we introduce two LQs with different representations of $SU(2)_L$ into the model. Thus, the correlation between $C_{9}$ and $C_{10}$ is diminished.  It is found that when the  $C_{10}$ is constrained by $B_s\to \mu^+ \mu^-$,  the  $C_9$ then can satisfy the requirements from the global analysis of $B\to K^* \mu^+ \mu^-$, and can also explain the anomaly of $R_K$, and the muon $g-2$ can fit the current data.


 The colored scalar LQs can couple to the SM Higgs  in the scalar potential; thus,  the LQ effects can influence the SM Higgs production and decays. The Higgs measurements have approached the precision level since the SM Higgs was discovered. Any sizable deviations from the SM predictions will indicate new physics.  In this study, we analyze  the  LQ-loop contributions to the diphoton Higgs decay.  It is worth mentioning that the introduced LQs can significantly  enhance the production cross section of a heavy scalar boson if such heavy scalar is probed at the LHC in the future.  The  relevant  studies on the heavy scalar production via LQ couplings  can be found in Refs.~\cite{DiIura:2016wbx,Dey:2016sht,Deppisch:2016qqd,Hati:2016thk,Chao:2015nac,Murphy:2015kag,Bauer:2015boy}. 

The paper is organized as follows. We introduce the model and discuss the relevant couplings in Sec.~II. In Sec.~III, we study the phenomena: the SM Higgs diphoton decay, LFV processes,  Wilson coefficients of $C_{9,10}$ from LQ contributions for $b\to s\ell^+ \ell^-$ decays, and the implication of $B_s \to \mu^+ \mu^-$. The conclusion is given in Sec. IV.

\section{ Couplings to the leptoquarks}

In this section, we briefly introduce  the model and relevant interactions with the LQs. To reconcile the measurements of $B_s\to \mu ^+ \mu^-$ and $B\to K^* \ell^+ \ell^-$, we extend the SM by adding two different representations of LQ, which are  $\Phi_{7/6}=(3,2)_{7/6}$  and $\Delta_{1/3}=(\bar 3,3)_{1/3}$  under $(SU(3)_C, SU(2)_L)_{U(1)_Y}$ SM gauge symmetry. 
The  gauge-invariant Yukawa interactions of the SM fermions and LQs are written as: 
\begin{align}
\label{eq:int1}
L_{LQ} =  k_{ij} \bar{Q}_i \Phi_{7/6 } \ell_{Rj} + \tilde{k}_{ij} \bar L_i \tilde \Phi_{7/6} u_{Rj} 
+ y_{ij} \bar{Q}_i^c i \sigma_2 \Delta_{1/3}  L_j + h.c.\,, 
\end{align}
where the subscripts $i\,, j$ are the flavor indices; $L^T_i = (\nu_i, \ell_i^-)$ and $Q^T_i = (u_i, d_i)$ are the $SU(2)_L$ lepton and quark doublets, $\tilde \Phi_{7/6} =i \sigma_2 \Phi^*_{7/6}$, and $(k_{ij}, \tilde k_{ij}, y_{ij})$ are the Yukawa couplings. Since we do not study the CP violating effects, hereafter,  we take all new Yukawa couplings as real numbers.
We use the representations of the LQs as:
\begin{align}
 \Phi_{7/6} = \begin{pmatrix} \phi^{5/3} \\ \phi^{2/3} \end{pmatrix}\,, \
\Delta_{1/3} = \begin{pmatrix} \delta^{1/3}/\sqrt{2} & \delta^{4/3} \\ \delta^{-2/3} & - \delta^{1/3}/\sqrt{2} \end{pmatrix}\,,
\end{align}
where the superscripts are the electric charges of the particles. The interactions in Eq.~(\ref{eq:int1}) are then expressed as:
\begin{align}
\label{eq:int2}
L_{LQ} =& k_{ij} \left[ \bar u_{Li} \, \ell_{Rj} \phi^{5/3}  + \bar d_{Li} \, \ell_{R j} \phi^{2/3}  \right] +  \tilde{k}_{ij} \left[ \bar \ell_{L i}  \, u_{Rj} \phi^{-5/3} - \bar \nu_{L i}\, u_{R j} \phi^{-2/3} \right]   \nonumber \\ 
&+ y_{ij} \left[ \bar u^c_{Li}  \, \nu_{Lj} \delta^{-2/3} - \frac{1}{\sqrt{2}} \bar u^c_{L i}  \, \ell_{L j} \delta^{1/3} - \frac{1}{\sqrt{2}} \bar d^c_{L i} \, \nu_{L j} \delta^{1/3} - \bar d^c_{L i}  \, \ell_{L j} \delta^{4/3} \right] + h.c. 
\end{align}

Since the LQs are colored scalar bosons, they can couple to the SM Higgs $H$ via the scalar potential. In order to get the Higgs couplings to the LQs, we write the gauge-invariant scalar potential as:
 \begin{align}
\label{eq:potential}
V & =  \mu^2  H^\dagger H + \lambda (H^\dagger H)^2+  M_\Phi^2 \left(\Phi_{7/6}^\dagger \Phi_{7/6} \right) 
 + M_{\Delta}^2 Tr \left( \Delta_{1/3}^\dagger \Delta_{1/3} \right) + \lambda_\Phi \left(\Phi_{7/6}^\dagger \Phi_{7/6} \right)^2   \nonumber \\
 & + \lambda_\Delta \left [Tr \left( \Delta_{1/3}^\dagger \Delta_{1/3}\right)  \right]^2
+ \lambda'_\Delta Tr \left( \left[ \Delta_{1/3}^\dagger \Delta_{1/3} \right]^2 \right)  + \lambda_{H \Phi} (H^\dagger H)\left(\Phi_{7/6}^\dagger \Phi_{7/6} \right) \nonumber \\
&+ \lambda_{H \Delta}  (H^\dagger H) Tr\left( \Delta_{1/3}^\dagger \Delta_{1/3} \right)
+ \lambda_{\Phi \Delta} \left(\Phi_{7/6}^\dagger \Phi_{7/6} \right)  Tr \left( \Delta_{1/3}^\dagger \Delta_{1/3} \right)\,.
 \end{align}
As usual, we adopt the representations of the Higgs doublet $H$ as:
\begin{equation}
H = \begin{pmatrix} G^+ \\ \frac{1}{\sqrt{2}} ( v+ \phi + iG^0) \end{pmatrix}\,, 
\end{equation}
where $G^+$ and $G^0$ are the Goldstone bosons; $\phi$ is the SM Higgs field,  and $v$ is the vacuum expectation value (VEV) of $H$. It is known that the VEV of scalar field is dictated by the scalar potential.


\section{Phenomenological Analysis}

Based on the introduced new interactions, in this section, we study the implications of the Higgs diphoton decay,  $\ell_i \to \ell_j \gamma$, the muon $g-2$, $h\to \tau \mu$, $B\to K^* \ell^+ \ell^-$, and $R_K$. Since each of these processes has its own unique characteristics, we discuss these phenomena one by one below.  

\subsection{Higgs diphoton decay}

The Higgs measurement  is usually described by the signal strength parameter, which is defined as the ratio of observation to the SM prediction and expressed as:
  \be
  \mu^f_i =\frac{\sigma(pp\to h) }{\sigma(pp\to h)_{\rm SM} } \cdot \frac{ \text{BR}(h\to f)}{ \text{BR}(h\to f)_{\rm SM}} \equiv \mu_i \cdot \mu_f \,, \label{eq:muf}
  \ee
 where $f$ stands for the possible channels, and $\mu_i (\mu_f)$ denotes the signal strength of production (decay). Although vector-boson fusion (VBF) can also produce the SM Higgs, we only consider the gluon-gluon fusion (ggF) process because it is the most dominant.  The diphoton Higgs decay approached  the precision measurement since the 125 GeV Higgs  was observed.  Therefore, any significant deviation from the SM prediction (i.e., $\mu^f_i \neq 1$) can imply the new physical effects.

As stated earlier, the SM Higgs can couple to the LQs via the scalar potential. From Eq.~(\ref{eq:potential}), it can be seen that after spontaneous symmetry breaking (SSB), the quartic terms  $H^\dagger H\Phi^\dagger_{7/6} \Phi_{7/6}$ and $H^\dagger H Tr(\Delta^\dagger_{1/3} \Delta_{1/3})$ can lead to trilinear couplings of Higgs to LQs as:
\begin{align}
{\cal L} & \supset 
\mu_{h \Phi} h \left(\phi^{-5/3} \phi^{5/3} + \phi^{-2/3} \phi^{2/3} \right) + \mu_{h \Delta}h \left( \delta^{-1/3} \delta^{1/3} + \delta^{-2/3} \delta^{2/3} + \delta^{-4/3} \delta^{4/3} \right)\,, \label{eq:hLQ}
\end{align}
where $\mu_{h\Phi} = \lambda_{H\Phi} v$ and $\mu_{h\Delta} = \lambda_{H\Delta} v$. 
With the couplings in Eq.~(\ref{eq:hLQ}),  the effective Lagrangian for $hgg$ by LQ-loop can be formulated as:
\begin{equation}
\Delta {\cal L}_{hgg} = \frac{\alpha_s}{8 \pi} \left( \frac{ \mu_{h \Phi} }{m_\Phi^2 } A_0(\xi_\Phi) + \frac{3 \mu_{h \Delta}}{2 m_\Delta^2} A_0(\xi_\Delta) \right) h G^{a \mu \nu} G^a_{\mu \nu}\,, \label{eq:hgg}
\end{equation}
where $\xi_X = 4 m_{X}^2/m_h^2$  and the loop function is given by:
\begin{equation}
A_0 (x) = x (1 -x f(x))
\end{equation}
with $f(x) = \left[\sin^{-1} (1/\sqrt{x})\right]^2 $ for $x > 1$. Accordingly, the signal strength of the Higgs production and decay to diphoton can be respectively obtained  as:
\begin{align}
\mu_i & = \left| 1 +  \frac{v}{A_{1/2}(\xi_t)}\sum_{X=\Phi,\Delta} \frac{n_X \mu_{hX} }{m^2_X}A_0(\xi_X)
  \right|^2\,, \nonumber \\
\mu_{\gamma \gamma }  & = \left| 1 + \frac{v N_c}{2}  
\frac{\sum_{X=\Phi, \Delta} Q^2_X  A_0(\xi_X) \mu_{hX}/m^2_X}
{A_1(\xi_W) + Q^2_t N_c A_{1/2}(\xi_t)} \right|^2,
\end{align}
where $n_{\Phi(\Delta)}=2(3)$, $N_c=3$ is the number of colors; $Q^2_\Phi = 29/9$ and $Q^2_\Delta = 21/9$, and the functions for vector-boson and fermion loops are given by
\begin{align}
 A_{1/2}(x) & = -2[x + (1-x) f(x))]\,, \nonumber  \\
 A_1(x)  & = 2 + 3 x + 3(2x -x^2)f(x)\,.
\end{align}

\begin{figure}[hpbt] 
\includegraphics[width=70mm]{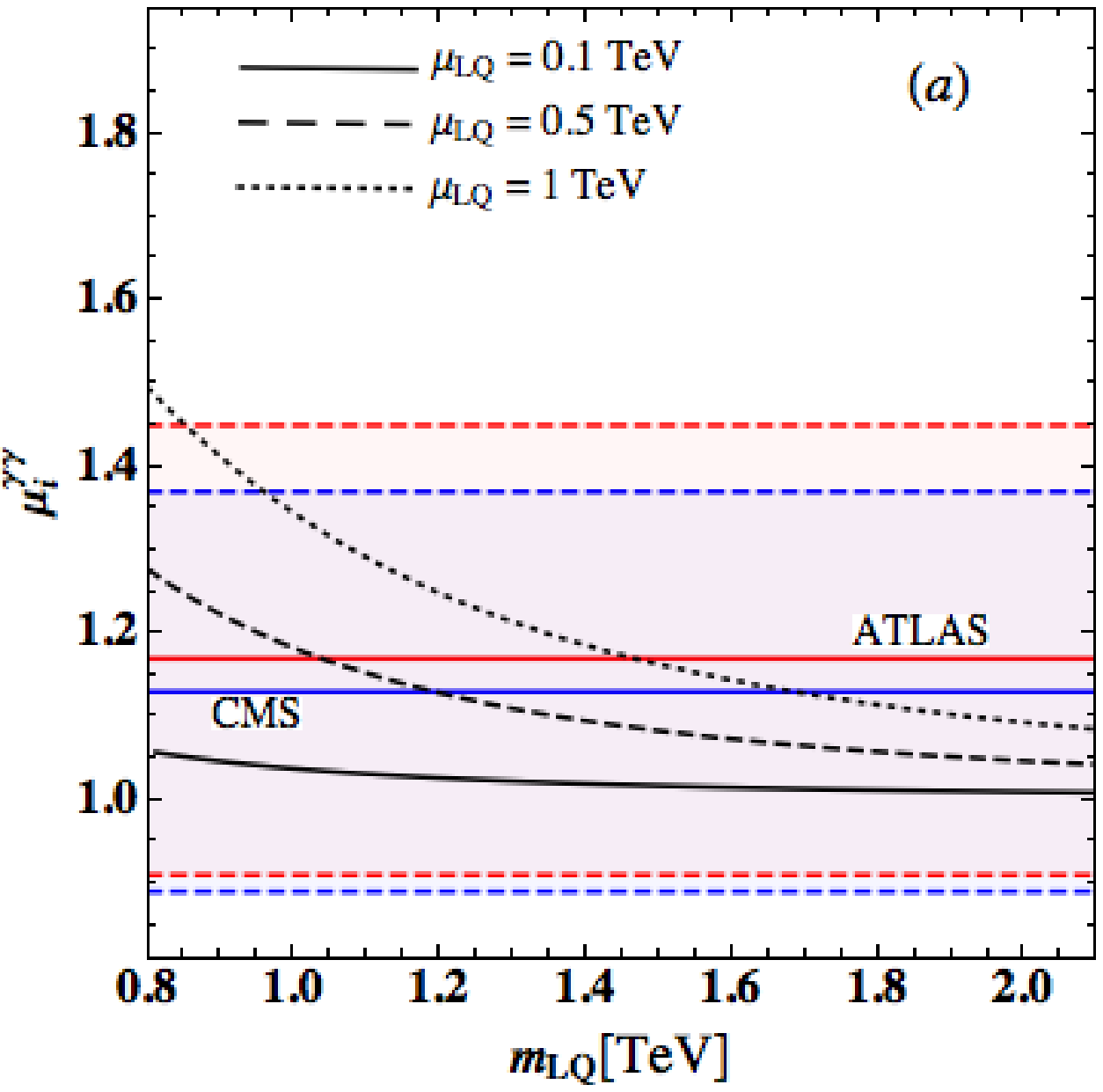} 
\includegraphics[width=70mm]{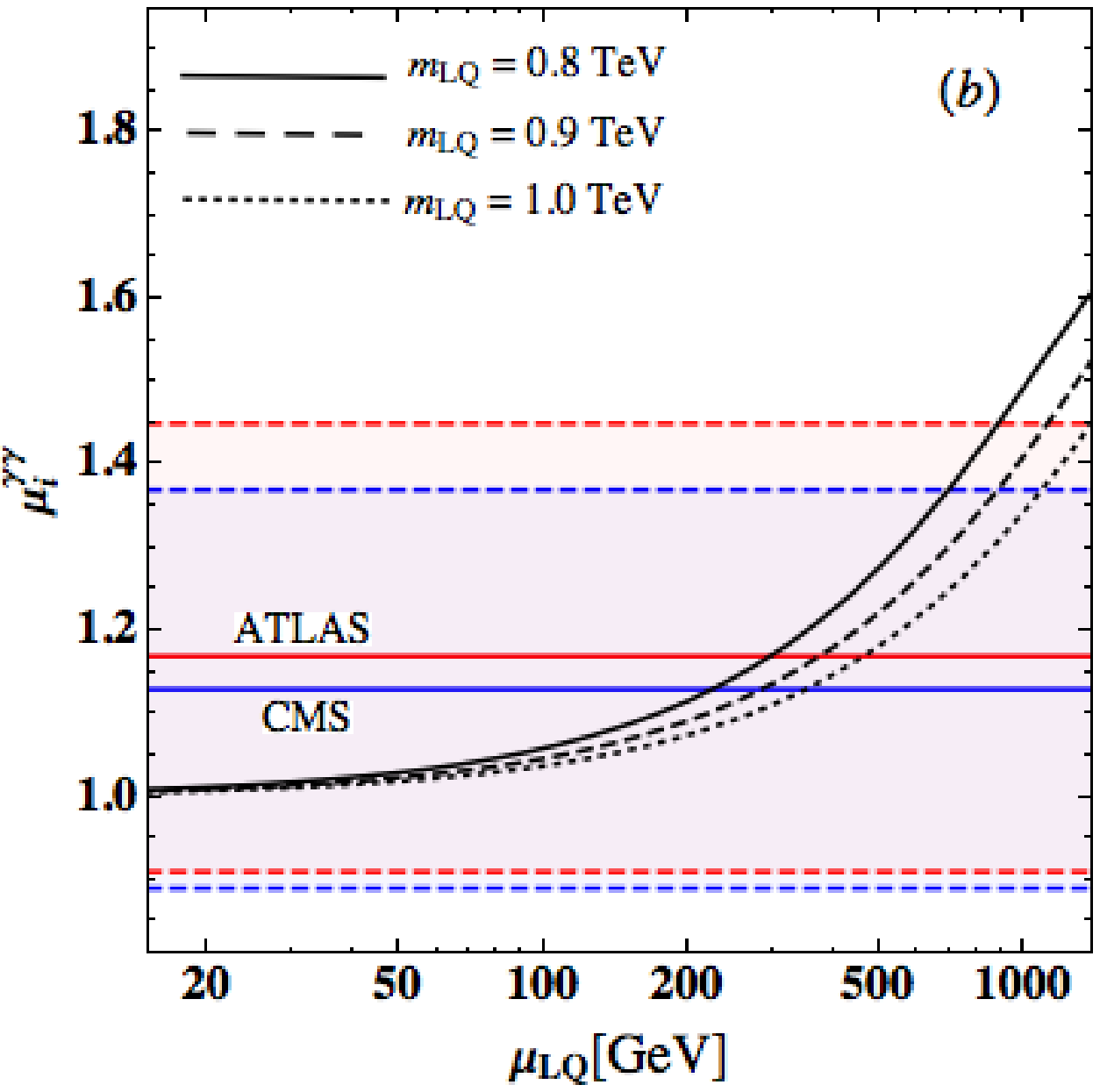} 
%
\caption{ Diphoton signal strength parameter $\mu^{\gamma\gamma}_i$ as a function of (a) $m_{LQ}$ and (b) $\mu_{LQ}$, where the curves in plots (a) and (b) denote $\mu_{LQ}=(0.1, 0.5,1)$ TeV and $m_{LQ}=(0.8,0.9,1.0)$ TeV, respectively. }
\label{fig:muif}
\end{figure}

Since the effects of the doublet and triplet LQs are similar, for simplicity, we set $\mu_{h\Phi}=\mu_{h \Delta} = \mu_{LQ}$ and $m_\Phi = m_\Delta = m_{LQ}$. The $\mu^{\gamma\gamma}_i$ as a function of $m_{LQ}$ is presented  in Fig.~\ref{fig:muif}(a), and that of $\mu_{LQ}$ is shown  in Fig.~\ref{fig:muif}(b), 
where the curves in plot (a) are $\mu_{LQ}=0.1,0.5,1$ TeV, and those in plot (b) are $m_{LQ}=0.8, 0.9, 1.0$ TeV.
For comparison,  we also show the results of  ATLAS~\cite{Aad:2015gba} and CMS~\cite{CMS:2014ega}  with $1\sigma$ errors  in the plots. From the plots, it can be clearly  seen that with $\mu_{LQ}$ of ${\cal O}(100)$ GeV, the LQ contributions can significantly shift the $\mu^{\gamma\gamma}_i$ away from the SM prediction and that the results are consistent with the current data. On the contrary, the $\mu^{\gamma\gamma}_i$  approaches the SM result when $\mu_{LQ}$ is of the order of GeV.

\subsection{ Radiative and Higgs LFV processes, muon $g-2$, and $B \to K^{(*)} \ell^+ \ell'^-$ decays}

In the following analysis, we study the rare
 lepton-flavor violating processes, e.g. $\ell_i \to \ell_j \gamma$ and $h\to \bar\tau  \mu + \bar\mu \tau$, muon $g-2$
 $\Delta a_\mu$, and the FCNC process $B\to K^* \ell^+ \ell^-$. We first discuss the radiative LFV processes for $\ell_i \to \ell_j \gamma$.  With the couplings in Eq.~(\ref{eq:int2}), the LQ-loop induced decay amplitude for $\ell_i \to \ell_j \gamma$ can be written as:
 \begin{eqnarray}
 {\cal L}_{\ell_i \to \ell_j \gamma}  = \frac{e}{2} \bar \ell_j \sigma_{\mu\nu} \left[ (c_L)_{ji} P_L + (c_R)_{ji} P_R  \right] \ell_i F^{\mu \nu} \,,\label{eq:Lllga}
 \end{eqnarray} 
where the coefficient $(c_{R})_{ji}$ is expressed as:
\begin{align}
(c_R)_{ji} & \approx   \frac{m_t}{(4 \pi)^2} ( k^\dagger)_{i 3} \tilde k_{3 j} \int d[X]    \left(  \frac{5}{\Delta(m_{t}, m_\Phi) } - \frac{2(1-x)}{\Delta(m_\Phi, m_{t}) }\right) \,, \nonumber \\
%
\Delta(m_1,m_2)  & =    x m_1^2 +(y+z) m_2^2 \,, \nonumber \\
\int [dX] & =  \int dx dy dz \delta(1-x-y-z)\,, \label{eq:cRji}
\end{align}
$(c_L)_{ji}$ can be obtained from $(c_R)_{ji}$ by exchanging $k_{ab}$ and  $\tilde k_{ab}$. In order to  balance the chirality of the leptons, it is found that the contributions from $k^\dagger_{iq} k_{q j}$, $\tilde{k}^\dagger_{iq} \tilde{k}_{q j}$, $y^\dagger_{iq} k_{qj}$, and $y^\dagger_{iq} y_{qj}$ are suppressed by the lepton masses. Since  the LQ $\phi^{5/3}$ can couple to left-handed and right-handed up-type quarks, the chirality flip by the mass insertion in the propagator of the up-type quark  can lead to  freeing  of the lepton masses  in the Feynman diagrams, which  are associated with $k_{qi}$ and $\tilde k_{qi}$. In addition, the top-quark is much heavier than the $u$- and $c$-quarks; therefore, we only present the top-quark contribution in $(c_R)_{ji}$. Straightforwardly, the BR for $\ell_i \to \ell_j \gamma$ can be expressed as:
\begin{equation}
\text{BR}(\ell_i \to \ell_j \gamma) = \frac{48 \pi^3 \alpha \eta_i}{G_F^2 m_{\ell_i}^2} \left( \left| (c_R)_{ji} \right|^2 + \left| (c_L)_{ji} \right|^2 \right)\,, 
\end{equation}
where $\eta_i \simeq (1,1/5)$ for $i = (\mu, \tau)$ and the BRs for $\ell_i \to \ell_j \bar \nu_j \nu_i$ in the SM have been applied. The current experimental upper limits are shown in Table~\ref{tab:Cif}. 
%
According to Eq.~(\ref{eq:Lllga}), muon $g-2$ 
can be easily obtained by setting $j=i=\mu$ and found as:
\begin{equation}
\Delta a_\mu \simeq - \frac{m_\mu}{2} (c_L + c_R)_{\mu\mu}. \label{eq:mug2}
\end{equation}

\begin{table}[hpbt]
\begin{tabular}{c|c|c} \hline
Process & $(i,j)$ & Experimental bounds ($90\%$ CL) \\ \hline
$\mu^{-} \to e^{-} \gamma$ & $(2,1)$ &
	$\text{BR}(\mu \to e\gamma) < 5.7 \times 10^{-13}$  \\
$\tau^{-} \to e^{-} \gamma$ & $(3,1)$ &
	$\text{BR}(\tau \to e\gamma) < 3.3 \times 10^{-8}$ \\
$\tau^{-} \to \mu^{-} \gamma$ & $(3,2)$ &
	$\text{BR}(\tau \to \mu\gamma) < 4.4 \times 10^{-8}$  \\ \hline
\end{tabular}
\caption{ Current upper bounds on the BRs for the decays $\ell_i \to \ell_j \gamma$~\cite{Adam:2013mnn}.}
\label{tab:Cif}
\end{table}

If   the photon in $\ell_i \to \ell_j \gamma$ is replaced by the Higgs, similar Feynman diagrams can  contribute to $h\to \bar \ell_j \ell_i + \bar \ell_i \ell_j \equiv \ell_j \ell_i$. Since  the upper limit  of $\text{BR}(\mu\to e\gamma)$ is of ${\cal O}(10^{-13})$ and can give  strong constraints on the parameters $k^\dagger_{23}\tilde k_{31}$ and $\tilde{k}^\dagger_{23} k_{31}$, if we set $k_{31}$ and $\tilde k_{31}$ to be small, then it is apparent that $h\to e \mu$ and $h\to e \tau$ are much smaller than current upper limits. Hence, we just study the decay $h\to \mu \tau$.
The one-loop induced effective couplings for $h \mu \tau$ are written as:
\begin{equation}
\mathcal{L} = h \bar \mu (C_R P_R + C_L P_L) \tau + h.c.\,,
\end{equation}
where $C_{L}$ is expressed as~\cite{Baek:2015mea,Baek:2016kud}:
\begin{align}
C_L =& \frac{(k^\dagger)_{23} \tilde k_{33} N_c m_t}{(4 \pi)^2 v} \left[ A\left(\frac{m^2_t}{m^2_\Phi}, \frac{m^2_h}{m^2_\Phi} \right)  + B\left(\frac{m^2_t}{m^2_\Phi}, \frac{m^2_h}{m^2_\Phi} \right)  \right] \nonumber \\
& +  \frac{ N_c \mu_{h \Phi} }{(4\pi)^2} \sum_{i=1\text{-}3, q=u,d} \left[ m_\mu (k^\dagger)_{2 i} k_{i 3} G(m_{q_i},m_\Phi )+ m_\tau \tilde k^\dagger_{2 i} \tilde k_{i 3} \tilde G(m_{q_i},m_\Phi) \right]  \nonumber \\
& + \frac{N_c m_\tau \mu_{h \Delta}}{(4 \pi)^2} \sum_{i=1\text{-}3,q=u,d} (y^\dagger)_{2 i} y_{i 3}  \tilde G(m_{q_i},m_\Delta) \,, 
\label{eq:CL}
\end{align}
$C_R$ can be obtained from $C_L$ by exchanging $k_{ab}$ and $\tilde k_{ab}$, and
the loop functions are given by:
\begin{align}
A(r_t, r_h) =& -\frac{1}{2} -2 \int [dX] \log \left[z+(1-z)r_t - x y r_h - i \epsilon \right] \nonumber \\
& + \int_0^1 dx \log \left[ x + (1-x) r_t - i \epsilon \right]\,, \nonumber  \\
B(r_t, r_h) = & \int [dX] \frac{xy r_h - r_t }{z+(1-z)r_t - xy r_h} \,, \nonumber \\
G(m_1,m_2) \approx  & \int [dX] \frac{ z}{- x z m_h^2 + x m_1^2 + (y+z) m_2^2 }\,, \nonumber \\
\tilde G(m_1,m_2) \approx & \int [dX] \frac{ y}{ - x z m_h^2 + x m_1^2 + (y+z) m_2^2 } \,.
\end{align}
 The $\epsilon$ in $A(r_t, r_h)$ denotes an infinitesimal positive value.
It can be seen that the terms associated with  $k^\dagger_{2i} k_{i3}$, $\tilde{k}^\dagger_{2i} k_{i3}$, and $y^\dagger_{2i} y_{i3}$ in Eq.~(\ref{eq:CL}) are proportional to the lepton masses. The situation is similar to the $(c_R)_{ji}$ in the decays  $\ell_i \to \ell_j \gamma$. Although $\mu_{hX}$  of  TeV ($X=\Phi, \Delta)$ can enhance these effects, due to the effects of being related to  $\mu_{hX} m_\ell /m^2_X$, their contributions are at least $10^{-2}$ smaller than those from $k^\dagger_{2i} \tilde{k}_{i3}$. Accordingly, the BR for $h \to \mu \tau$ is formulated as:
\begin{align}
& \text{BR}(h\to  \mu \tau)  \approx  \frac{m_h}{16\pi \Gamma_h}  (|C_L|^2 + | C_R|^2)\,,
\end{align}
where $\Gamma_h$ is the width of the Higgs. Due to $\text{BR}(h\to  \mu \tau)$ being less than $1\%$, we use $\Gamma_h\approx \Gamma^{\rm SM}_h \approx 4.2$ MeV in our numerical estimations.  


Next, we discuss the decays for $b\to s \ell^+ \ell^-$. 
In order to include the effects of lepton non-unversality, we write the effective Hamiltonian as:
 \begin{equation}
 {\cal H}  = \frac{G_F \alpha V_{tb} V^*_{ts}}{\sqrt{2}\pi} \left[ H_{1\mu} L^\mu + H_{2\mu} L^{5\mu} \right]\,, \label{eq:Hbsll}
 \end{equation}
where the leptonic currents are denoted by $L^{(5)}_\mu= \ell \gamma_\mu (\gamma_5) \ell$; and  the related hadronic currents are defined as:
 \begin{align}
 H_{1\mu} &= C^\ell_9 \bar s \gamma_\mu P_L - \frac{2m_b}{q^2} C_7 \bar s i \sigma_{\mu \nu} q^\nu P_R b\,, \nonumber \\
 H_{2\mu} & = C^{\ell}_{10} \bar s \gamma_\mu P_L b\,.
 \end{align}
Here, the Wilson coefficients are read as: $C^\ell_{9(10)} = C^{\rm SM}_{9(10)} + C^{\rm NP,\ell}_{9(10)}$ and $C_7 = C^{\rm SM}_{7}$. The detailed angular distribution for $B\to (K\pi)_{K^*} \ell^+ \ell^-$ can be found in Refs.~\cite{DescotesGenon:2012zf,Chen:2002bq,Chen:2002zk,Altmannshofer:2008dz,Egede:2010zc}.  Following the notations in Ref.~\cite{DescotesGenon:2012zf}, the angular observable $P'_5$ is defined by:
\begin{align}
P'_5 &= \frac{J_5}{ \sqrt{-J_{2c} J_{2s} }} \,, \quad J_5 = \sqrt{2} Re(A^L_0 A^{L*}_{\perp} )\,,  \nonumber \\
 J_{2c} &=- |A^L_0|^2 \,, \quad J_{2s} =\frac{1}{4} \left(|A^L_{\parallel}|^2 + |A^L_{\perp}|^2 \right) \,,
\end{align}
where $A^{L}_{0,\parallel,\perp}$ are related to the $B\to K^* $ transition form factors and the Wilson coefficients of $C^\ell_{9, 10}$ and $C_7$. Their explicit expressions can be found in Ref.~\cite{DescotesGenon:2012zf}. In this study, we do not directly investigate the angular analysis of $B\to K^* \ell^+ \ell^-$; instead, we  refer to the results, which were done by using the  global analysis to get the best-fit value of $C^{NP}_9 \approx -1.09$ for the new physics contributions~\cite{Descotes-Genon:2015uva}.   Thus, we just derive the Wilson coefficients of $C^\ell_9$ and $C^\ell_{10}$ from the LQ contributions.  

With the Yukawa couplings in Eq.~(\ref{eq:int2}), the effective Hamiltonian for $b\to s \ell^+ \ell^-$  mediated by $\phi^{2/3}$ and $\delta^{4/3}$ can be respectively found as:
\begin{align}
& H_{\rm eff}^1 = \frac{k_{b \ell} k_{s \ell}}{2 m^2_{\Phi}} (\bar{s}\gamma^\mu P_L b)(\bar{\ell} \gamma_\mu P_{ R} \ell)\,, \nonumber
 \\
& H_{\rm eff}^2 = -\frac{y_{b \ell} y_{s \ell}}{2 m^2_{\Delta}} (\bar{s}\gamma^\mu P_L b)(\bar{\ell} \gamma_\mu P_{L} \ell)\,. \label{eq:Heff}
\end{align}
%
%
We can decompose  the Eq.~(\ref{eq:Heff}) in terms of  the effective operators $O_9$ and $O_{10}$, defined as
 $O_{9(10)} = \bar s  \gamma_\mu P_L b \; \bar \ell \gamma^\mu (\gamma_5) \ell $. The associated Wilson coefficients of $O_{9,10}$ from the LQs then are found as:
\begin{align}
C^{LQ,\ell}_9 &= -  \frac{1}{c_{\rm SM}} \left(\frac{k_{b \ell} k_{s \ell}}{4 m^2_{\Phi}} -\frac{y_{b \ell} y_{s \ell}}{4 m^2_{\Delta}} \right)\,, \nonumber  \\
C^{LQ,\ell}_{10} &= \frac{1}{c_{\rm SM}} 
 \left(\frac{k_{b \ell} k_{s \ell}}{4 m^2_{\Phi}} + \frac{y_{b \ell} y_{s \ell}}{4 m^2_{\Delta}} \right)\,, \label{eq:C9C10}
\end{align}
where $c_{\rm SM} = V_{tb} V^*_{ts} \alpha G_F/(\sqrt{2} \pi) $ is a scale  factor from the SM effective Hamiltonian. 
It is worth mentioning that the interaction $C^{LQ,\mu}_{10} O_{10}$ can contribute  to $B_s \to \mu^+ \mu^-$.
%
 Since the experimental data are consistent with the SM prediction, to consider the constraint from $B_s \to \mu^+ \mu^-$, we adopt the expression for the BR as~\cite{Hiller:2014yaa}:
 \begin{equation}
\frac{\text{BR}(B_s \to \mu^+ \mu^-)}{\text{BR}(B_s \to \mu^+ \mu^-)^{\text SM}} = \left|1-0.24 C^{LQ,\mu}_{10} \right|^2.
\end{equation}
With $1\sigma$ errors, the allowed range for $C^{LQ,\mu}_{10}$ is obtained as $C^{LQ,\mu}_{10}=(0.21, 0.79)$. We use this result to constrain the free parameters. Since the $R_K$ is insensitive to the $B\to K$ transition form factors~\cite{Hiller:2003js},  in order to study the anomaly of $R_K$, we require that  the allowed range of parameters  has to satisfy~\cite{Hiller:2014yaa}:
 \begin{align}
 0.7 \leq Re[X^e - X^\mu] \leq 1.5\,, \label{eq:X}
 \end{align}
where $X^\ell = C^{LQ,\ell}_9 - C^{LQ,\ell}_{10}$, and the $R_K$ data with $1\sigma$ errors are used.

%
%


Since the parameters in the decays $\ell_i \to \ell_j \gamma$, $h\to \mu \tau$, $\Delta a_\mu$, and $B\to K^{(*)} \ell^+ \ell^-$ are strongly correlated, in the following analysis, we take the current upper limits of $\text{BR}(\ell_i \to \ell_j \gamma)$ shown in Table~\ref{tab:Cif} as the inputs and attempt to find the allowed parameter space,  such that the excesses in  $\Delta a_\mu$ and $B\to K^{(*)} \ell^+ \ell^-$ can be satisfied, and the $\text{BR}(h\to \mu \tau)$ can be as large as possible.

 From $(c_R)_{ji}$ in Eq.~(\ref{eq:cRji}), the dominant effects on the radiative LFV processes are from the $\phi^{5/3}$ and the top-quark loop;  thus,  there is no  possible cancellation in any of the decay amplitudes. With the upper bound of $\text{BR}(\mu\to e \gamma)$, we see that $k^\dagger_{13} \tilde k_{32}$  and $\tilde k^\dagger_{13}  k_{32}$ have to be very small. In order to explain the excesses of muon $g-2$ and $B\to K^* \mu^+ \mu^-$, we set $k_{31}=\tilde k_{31}\approx  0$.  As a result, $\text{BR}(h\to e \mu)$ is negligible in this model. The related parameters  for $\tau \to (\mu ,e) \gamma$ decays are $k_{31,32}\tilde k_{33}$ and $\tilde k_{31,32} k_{33}$, respectively. These parameters simultaneously influence $h\to (\mu, e) \tau$, muon $g-2$, 
 and $b\to s \mu^+ \mu^-$; therefore we have to  analyze  these processes together to get the allowed parameter space. 

Since Eqs.~(\ref{eq:cRji}), (\ref{eq:CL}), and (\ref{eq:C9C10}) involve many free parameters, in order to efficiently perform a numerical analysis, we set the ranges of relevant parameters as:
\begin{align}
& m_{LQ} \in [700\,, 1500\,] \; \text{GeV},\quad  \mu_{LQ} \in [1\,, 100]\; \text{GeV}\,, \quad \{ k_{22}, \tilde k_{22}, y_{22} \}  \in [-0.1\,, 0.1]\,,\nonumber \\
%
&  \{ k_{33},\tilde k_{33}, y_{33} \}  \in [-0.01,0.01]\,, \quad \{ k_{23},\tilde k_{23}, y_{23} \}  \in [-0.1\,, 0.1]\,,  \nonumber \\
& k_{32} \in  \text{sign}(k_{22}) [0,0.5 ]\,, \quad \tilde k_{32} \in [-0.5\,, 0.5]\,, \quad y_{32} \in - \text{sign}(y_{22}) [0\,, 0.5 ] \,.
	\label{eq:range_scanning}
\end{align} 
In order to avoid the constraints from $\tau\to \ell \gamma$ ($\ell = e, \mu$) and get $|C^{LQ,\mu}_{9}| \sim 1$, we set ($k_{33}/k_{32}, \tilde k_{33}/\tilde k_{32}) \sim 0.1$ in Eq.~(\ref{eq:range_scanning}). Additionally, the negative value of $C^{LQ,\mu}_{9}$ can be achieved when $k_{32}(y_{32})$ and $k_{22}(y_{22})$ are opposite in sign. 
%
%
As mentioned earlier, the Yukawa couplings in both decays $\tau \to \ell \gamma$ and $h\to \ell \tau$ are the same,  we cannot remove the constraints from the radiative LFV processes in this model. The BRs for $h\to \ell \tau$ thus are of ${\cal }O(10^{-9})$ and much smaller than the current upper limits of ${\cal O}(10^{-4})$~\cite{Khachatryan:2015kon,Aad:2015gha}. One way to escape the constraint from $\tau \to \ell \gamma$ is to add a new  LQ~\cite{Baek:2015mea}. Since we focus on the excesses of muon $g-2$
and $B\to K^* \mu^+ \mu^-$, we leave the more complicated model for further study.  

\begin{figure}[hptb] 
\includegraphics[width=70mm]{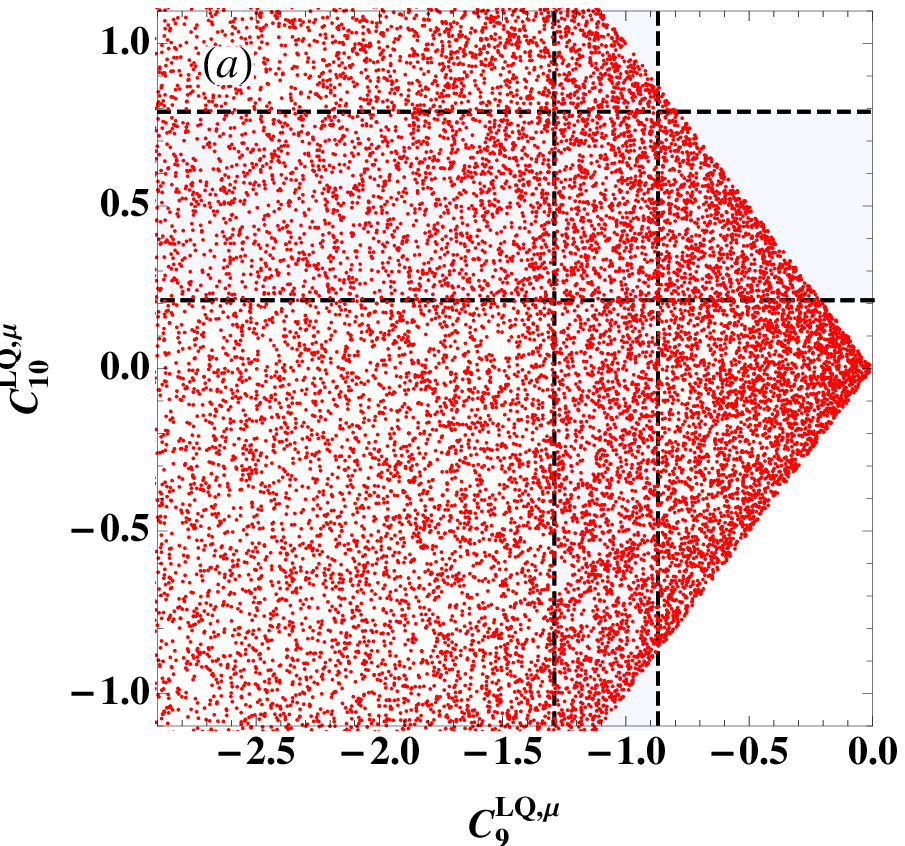} 
\includegraphics[width=70mm]{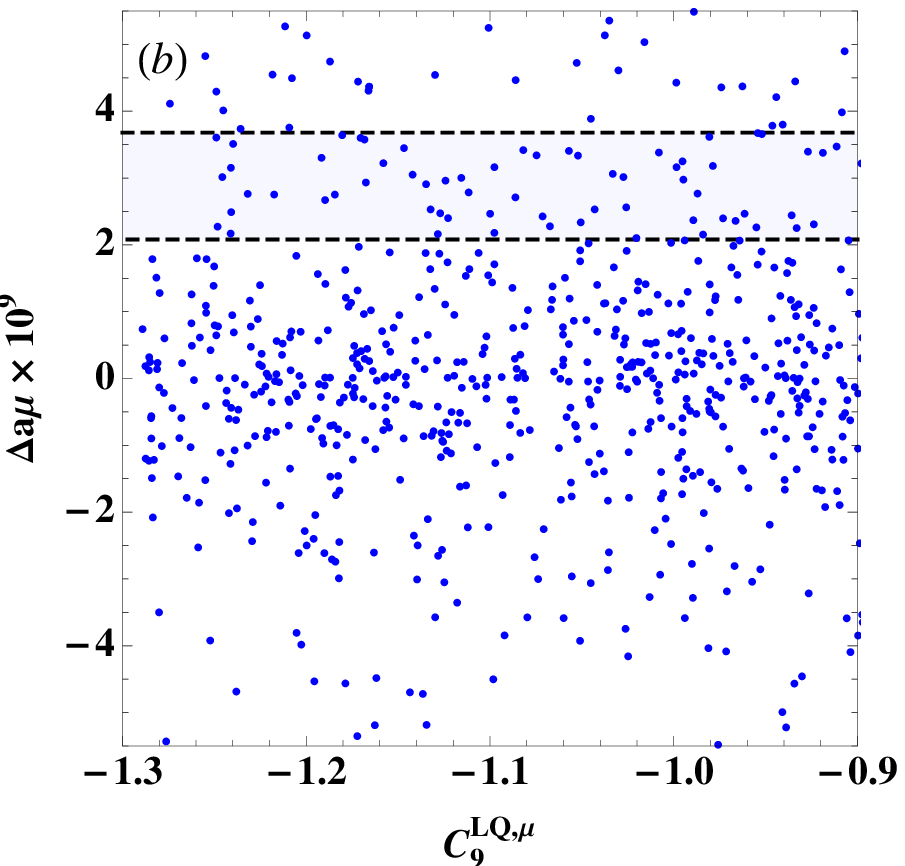} 
%
\caption{  (a) The values of $C^{LQ,\mu}_9$ and $C^{LQ,\mu}_{10}$ using the ranges of parameters  in Eq.~(\ref{eq:range_scanning}), where the bound from the $\text{BR}(B_s\to \mu^+ \mu^-)$ and the allowed range of $C^{LQ,\mu}_9$ from global analysis  for $B\to K^* \mu^+ \mu^-$ are shown. (b) Correlation of $\Delta a_\mu$ and $C^{LQ,\mu}_9$, where we only show the values of $C^{LQ,\mu}_9$ that can fit the excess in $B\to K^* \mu^+ \mu^-$, and the band bounded by two dashed lines denotes the $\Delta a_\mu$ data with $1 \sigma$ errors~\cite{PDG}. }
\label{fig:scatter}
\end{figure}

With the chosen ranges of parameters in Eq.~(\ref{eq:range_scanning}), we first show the values for $C^{LQ,\mu}_9$ and $C^{LQ,\mu}_{10}$ in Fig.~\ref{fig:scatter}(a), where   the bounds from $\tau \to \ell \gamma$ have been considered; the horizontal band is from the measurement  of $B_s\to \mu^+ \mu^-$;  the vertical band is the range that can explain the excess of $B\to K^* \mu^+ \mu^-$, and we  used  $10^{5}$ parameter sets and obtained  $824$ allowed points that satisfy the constraints. It can be seen that the $C^{LQ,\mu}_9$ and $C^{LQ,\mu}_{10}$ from the contributions of doublet $\Phi_{7/6}$ and triplet $\Delta_{1/3}$ LQs can simultaneously satisfy the constraint of $B_s\to \mu^+ \mu^-$ and explain the excess in $B\to K^* \mu^+ \mu^-$. 

 From Eq.~(\ref{eq:mug2}), it is known that muon $g-2$
 is associated with the Yukawa couplings $k_{32}\tilde k_{32}$. Although only $k_{32}$ is related to $C^{LQ,\mu}_9$ and $C^{LQ,\mu}_{10}$, since the Yukawa couplings $k_{q\ell}$,  $\tilde k_{q\ell}$, and $y_{q\ell}$ are taken to be the same order of magnitude, we present the correlations of $\Delta a_\mu$  and $C^{LQ,\mu}_{9}$ in Fig.~\ref{fig:scatter}(b), where only the allowed range of $C^{LQ,\mu}_{9}$ is shown, and the region between two dashed lines denotes the $\Delta a_\mu$ data with $1\sigma$ errors. By plot (b), it can be seen clearly that the excesses in $\Delta a_\mu$  and $B\to K^* \mu^+ \mu^-$ can be simultaneously fitted in the model.  

\begin{figure}[hptb] 
\includegraphics[width=75mm]{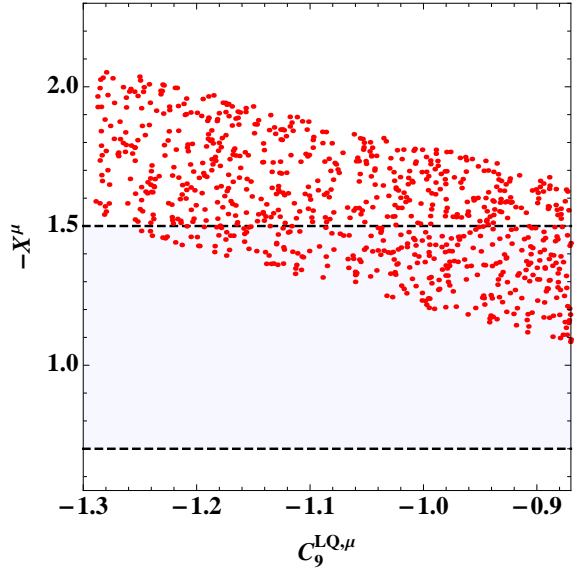} 
\caption{ Correlation between $X^\mu = C^{LQ,\mu}_9 - C^{LQ,\mu}_{10}$ and $C^{LQ,\mu}_9$, where the allowed range of $X^\mu$ is from the $R_K$ data with $1\sigma$ errors.
\label{fig:scatter2}}
\end{figure}

As discussed before, in order to avoid the constraint from $\mu\to e \gamma$, we set $k_{31}=\tilde k_{31}=0$ in our analysis; therefore, $C^{LQ,e}_{9(10)}$ for $B\to K e^+ e^-$ decay is only related to $y_{31} y_{21}$. Since $y_{31,21}$ are free parameters, for simplicity, we then take $|y_{31}|\sim |k_{31}|\sim 0$. As a result, $X^{e}=C^{LQ,e}_9 - C^{LQ,e}_{10} \approx 0$. In order to see whether the obtained $C^{LQ,\mu}_9$ and $C^{LQ,\mu}_{10}$ can fit the $R_K$ data, we show the correlation between $X^{\mu}$ and $C^{LQ,\mu}_{9}$ in 
Fig.~\ref{fig:scatter2}, where the band denotes the allowed range shown in Eq.~(\ref{eq:X}). It can be seen that the excesses of $R_K$ and $P'_5$ can be simultaneously explained when  the measurement of $B_s\to \mu^+ \mu^-$ is satisfied. 


\section{ Conclusion}

In order to resolve the excesses of muon $g-2$ and $B\to K^{(*)} \ell^+  \ell^-$ decays, we investigate the extension of the SM by  including leptoquarks, in which the particles are colored scalar bosons  and can couple to quarks and leptons.   In order to accommodate the measurement of $B_s\to \mu^+ \mu^-$ and the excesses of $B\to K^{(*)} \mu^+ \mu^-$, we study a model with  one doublet and  one triplet leptoquarks.

After SSB, the couplings of the SM Higgs the LQs are described by $\mu_{hX}=\lambda_{HX} v$. If $\mu_{hX}$ is of ${\cal O}(100)$ GeV, the signal strength parameter $\mu^{\gamma\gamma}_{i}$  can significantly deviate from the SM prediction and is still consistent with the current Higgs measurements. 

In this study, lepton-flavor violating processes $\ell_i \to \ell_j \gamma$ give strict constraints on the Yukawa couplings $k_{31,33}$ and  $\tilde k_{31,33}$. As a result, the branching ratios for the lepton-flavor violating Higgs decays $h\to \ell_i \ell_j$ are less than ${\cal O}(10^{-8}$). Nevertheless, the sizable couplings $k_{32,22}$, $\tilde k_{32,22}$, and $y_{32,22}$ can still explain the excess of muon $g-2$ and provide the necessary values for the Wilson coefficient $C^{\ell}_9$, such that the excesses in $B\to K^* \mu^+ \mu^-$ and $R_K$ can be resolved.



\section*{Acknowledgments}

This work was partially supported by the Ministry of Science and Technology of Taiwan
R.O.C.,  under grant MOST-103-2112-M-006-004-MY3 (CHC). H. O. would like to thank the members of KIAS for their hospitality  during his visit.


\begin{thebibliography}{99}

\bibitem{PDG} K.A. Olive et al. (Particle Data Group), Chin. Phys. C, 38, 090001 (2014).   


  \bibitem{DescotesGenon:2012zf} 
  S.~Descotes-Genon, J.~Matias, M.~Ramon and J.~Virto,
  JHEP {\bf 1301}, 048 (2013)
  [arXiv:1207.2753 [hep-ph]].


\bibitem{Aaij:2015oid} 
  R.~Aaij {\it et al.} [LHCb Collaboration],
  JHEP {\bf 1602}, 104 (2016)
  [arXiv:1512.04442 [hep-ex]].


\bibitem{Aaij:2013qta} 
  R.~Aaij {\it et al.} [LHCb Collaboration],
  Phys.\ Rev.\ Lett.\  {\bf 111}, 191801 (2013)
  [arXiv:1308.1707 [hep-ex]].
  

  
\bibitem{Abdesselam:2016llu} 
  A.~Abdesselam {\it et al.} [Belle Collaboration],
  arXiv:1604.04042 [hep-ex].
  
  \bibitem{Matias:2012xw} 
  J.~Matias, F.~Mescia, M.~Ramon and J.~Virto,
  JHEP {\bf 1204}, 104 (2012)
  [arXiv:1202.4266 [hep-ph]].

\bibitem{Descotes-Genon:2013wba} 
  S.~Descotes-Genon, J.~Matias and J.~Virto,
  Phys.\ Rev.\ D {\bf 88}, 074002 (2013)
  [arXiv:1307.5683 [hep-ph]].

  
\bibitem{Gauld:2013qja} 
  R.~Gauld, F.~Goertz and U.~Haisch,
  JHEP {\bf 1401}, 069 (2014)
  [arXiv:1310.1082 [hep-ph]].

\bibitem{Datta:2013kja} 
  A.~Datta, M.~Duraisamy and D.~Ghosh,
  Phys.\ Rev.\ D {\bf 89}, no. 7, 071501 (2014)
  [arXiv:1310.1937 [hep-ph]].

\bibitem{Hurth:2013ssa} 
  T.~Hurth and F.~Mahmoudi,
  JHEP {\bf 1404}, 097 (2014)
  [arXiv:1312.5267 [hep-ph]].
  


\bibitem{Descotes-Genon:2014uoa} 
  S.~Descotes-Genon, L.~Hofer, J.~Matias and J.~Virto,
  JHEP {\bf 1412}, 125 (2014)
  [arXiv:1407.8526 [hep-ph]].
 
\bibitem{Altmannshofer:2014rta} 
  W.~Altmannshofer and D.~M.~Straub,
  Eur.\ Phys.\ J.\ C {\bf 75}, no. 8, 382 (2015)
  [arXiv:1411.3161 [hep-ph]].
  

  
   
\bibitem{Descotes-Genon:2015uva} 
  S.~Descotes-Genon, L.~Hofer, J.~Matias and J.~Virto,
  JHEP {\bf 1606}, 092 (2016)
  [arXiv:1510.04239 [hep-ph]].
   
   
\bibitem{Crivellin:2015mga} 
  A.~Crivellin, G.~D'Ambrosio and J.~Heeck,
  Phys.\ Rev.\ Lett.\  {\bf 114}, 151801 (2015)
  [arXiv:1501.00993 [hep-ph]].

\bibitem{Sahoo:2015wya} 
  S.~Sahoo and R.~Mohanta,
  Phys.\ Rev.\ D {\bf 91}, no. 9, 094019 (2015)
  [arXiv:1501.05193 [hep-ph]].
  
\bibitem{Straub:2015ica} 
  A.~Bharucha, D.~M.~Straub and R.~Zwicky,
  arXiv:1503.05534 [hep-ph].
  
\bibitem{Becirevic:2015asa} 
  D.~Becirevic, S.~Fajfer and N.~Kosnik,
  Phys.\ Rev.\ D {\bf 92}, no. 1, 014016 (2015)
  [arXiv:1503.09024 [hep-ph]].

\bibitem{Crivellin:2015era} 
  A.~Crivellin, L.~Hofer, J.~Matias, U.~Nierste, S.~Pokorski and J.~Rosiek,
  Phys.\ Rev.\ D {\bf 92}, no. 5, 054013 (2015)
  [arXiv:1504.07928 [hep-ph]].
  
\bibitem{Lee:2015qra} 
  C.~J.~Lee and J.~Tandean,
  JHEP {\bf 1508}, 123 (2015)
  [arXiv:1505.04692 [hep-ph]].
  
\bibitem{Alonso:2015sja} 
  R.~Alonso, B.~Grinstein and J.~Martin Camalich,
  JHEP {\bf 1510}, 184 (2015)
  [arXiv:1505.05164 [hep-ph]].
  
\bibitem{Sahoo:2015qha} 
  S.~Sahoo and R.~Mohanta,
  Phys.\ Rev.\ D {\bf 93}, no. 3, 034018 (2016)
  [arXiv:1507.02070 [hep-ph]].
  
\bibitem{Belanger:2015nma} 
  G.~Belanger, C.~Delaunay and S.~Westhoff,
  Phys.\ Rev.\ D {\bf 92}, 055021 (2015)
  [arXiv:1507.06660 [hep-ph]].
  
\bibitem{Sahoo:2015pzk} 
  S.~Sahoo and R.~Mohanta,
  Phys.\ Rev.\ D {\bf 93}, no. 11, 114001 (2016)
  [arXiv:1512.04657 [hep-ph]].
  
\bibitem{Chiang:2016qov} 
  C.~W.~Chiang, X.~G.~He and G.~Valencia,
  Phys.\ Rev.\ D {\bf 93}, no. 7, 074003 (2016)
  [arXiv:1601.07328 [hep-ph]].
  
  
\bibitem{Dorsner:2016wpm} 
  I.~Dorsner, S.~Fajfer, A.~Greljo, J.~F.~Kamenik and N.~Kosnik,
  Phys.\ Rept.\  {\bf 641}, 1 (2016)
  [arXiv:1603.04993 [hep-ph]].
  
  \bibitem{Boucenna:2016wpr} 
  S.~M.~Boucenna, A.~Celis, J.~Fuentes-Martin, A.~Vicente and J.~Virto,
  Phys.\ Lett.\ B {\bf 760}, 214 (2016)
  [arXiv:1604.03088 [hep-ph]].
  

\bibitem{Hiller:2016kry} 
  G.~Hiller, D.~Loose and K.~Schonwald,
  arXiv:1609.08895 [hep-ph].


\bibitem{Aaij:2014ora} 
  R.~Aaij {\it et al.} [LHCb Collaboration],
  Phys.\ Rev.\ Lett.\  {\bf 113}, 151601 (2014)
  [arXiv:1406.6482 [hep-ex]].


\bibitem{Hiller:2014yaa} 
  G.~Hiller and M.~Schmaltz,
  Phys.\ Rev.\ D {\bf 90}, 054014 (2014)
  [arXiv:1408.1627 [hep-ph]].
  
  

\bibitem{Hurth:2014vma} 
  T.~Hurth, F.~Mahmoudi and S.~Neshatpour,
  JHEP {\bf 1412}, 053 (2014)
  [arXiv:1410.4545 [hep-ph]].
  

\bibitem{Glashow:2014iga} 
  S.~L.~Glashow, D.~Guadagnoli and K.~Lane,
  Phys.\ Rev.\ Lett.\  {\bf 114}, 091801 (2015)
  [arXiv:1411.0565 [hep-ph]].
  
\bibitem{Gripaios:2014tna} 
  B.~Gripaios, M.~Nardecchia and S.~A.~Renner,
  JHEP {\bf 1505}, 006 (2015)
  [arXiv:1412.1791 [hep-ph]].

  \bibitem{Sahoo:2015fla} 
  S.~Sahoo and R.~Mohanta,
  New J.\ Phys.\  {\bf 18}, no. 1, 013032 (2016)
  [arXiv:1509.06248 [hep-ph]].


\bibitem{Bauer:2015knc} 
  M.~Bauer and M.~Neubert,
  Phys.\ Rev.\ Lett.\  {\bf 116}, no. 14, 141802 (2016)
  [arXiv:1511.01900 [hep-ph]].
  
  
\bibitem{Das:2016vkr} 
  D.~Das, C.~Hati, G.~Kumar and N.~Mahajan,
  Phys.\ Rev.\ D {\bf 94}, no. 5, 055034 (2016)
  [arXiv:1605.06313 [hep-ph]].

\bibitem{Li:2016vvp} 
  X.~Q.~Li, Y.~D.~Yang and X.~Zhang,
  JHEP {\bf 1608}, 054 (2016)
  [arXiv:1605.09308 [hep-ph]].



\bibitem{Becirevic:2016yqi} 
  D.~Be$\breve{c}$irevi$\acute{c}$, S.~Fajfer, N.~Ko$\breve{s}$nik and O.~Sumensari,
  arXiv:1608.08501 [hep-ph].

\bibitem{Sahoo:2016pet} 
  S.~Sahoo, R.~Mohanta and A.~K.~Giri,
  arXiv:1609.04367 [hep-ph].


\bibitem{Bhattacharya:2016mcc} 
  B.~Bhattacharya, A.~Datta, J.~P.~Guevin, D.~London and R.~Watanabe,
  arXiv:1609.09078 [hep-ph].

\bibitem{Duraisamy:2016gsd} 
  M.~Duraisamy, S.~Sahoo and R.~Mohanta,
  arXiv:1610.00902 [hep-ph].
  

\bibitem{Aaboud:2016tru} 
  M.~Aaboud {\it et al.} [ATLAS Collaboration],
  arXiv:1606.03833 [hep-ex].


\bibitem{Khachatryan:2016hje} 
  V.~Khachatryan {\it et al.} [CMS Collaboration],
  arXiv:1606.04093 [hep-ex].
  
 

  


 
\bibitem{Harigaya:2015ezk} 
  K.~Harigaya and Y.~Nomura,
  Phys.\ Lett.\ B {\bf 754}, 151 (2016)
  [arXiv:1512.04850 [hep-ph]].
  
\bibitem{Mambrini:2015wyu} 
  Y.~Mambrini, G.~Arcadi and A.~Djouadi,
  Phys.\ Lett.\ B {\bf 755}, 426 (2016)
  [arXiv:1512.04913 [hep-ph]].
  
\bibitem{Backovic:2015fnp} 
  M.~Backovic, A.~Mariotti and D.~Redigolo,
  JHEP {\bf 1603}, 157 (2016)
  [arXiv:1512.04917 [hep-ph]].
  
\bibitem{Angelescu:2015uiz} 
  A.~Angelescu, A.~Djouadi and G.~Moreau,
  Phys.\ Lett.\ B {\bf 756}, 126 (2016)
  [arXiv:1512.04921 [hep-ph]].

 
\bibitem{Nakai:2015ptz} 
  Y.~Nakai, R.~Sato and K.~Tobioka,
  Phys.\ Rev.\ Lett.\  {\bf 116}, no. 15, 151802 (2016)
  [arXiv:1512.04924 [hep-ph]].
  
\bibitem{Buttazzo:2015txu} 
  D.~Buttazzo, A.~Greljo and D.~Marzocca,
  Eur.\ Phys.\ J.\ C {\bf 76}, no. 3, 116 (2016)
  [arXiv:1512.04929 [hep-ph]].
  
\bibitem{DiChiara:2015vdm} 
  S.~Di Chiara, L.~Marzola and M.~Raidal,
  Phys.\ Rev.\ D {\bf 93}, no. 9, 095018 (2016)
  [arXiv:1512.04939 [hep-ph]].
  
\bibitem{Knapen:2015dap} 
  S.~Knapen, T.~Melia, M.~Papucci and K.~Zurek,
  Phys.\ Rev.\ D {\bf 93}, no. 7, 075020 (2016)
  [arXiv:1512.04928 [hep-ph]].
  
\bibitem{Pilaftsis:2015ycr} 
  A.~Pilaftsis,
  Phys.\ Rev.\ D {\bf 93}, no. 1, 015017 (2016)
  [arXiv:1512.04931 [hep-ph]].
  
\bibitem{Franceschini:2015kwy} 
  R.~Franceschini {\it et al.},
  JHEP {\bf 1603}, 144 (2016)
  [arXiv:1512.04933 [hep-ph]].
  
\bibitem{Ellis:2015oso} 
  J.~Ellis, S.~A.~R.~Ellis, J.~Quevillon, V.~Sanz and T.~You,
  JHEP {\bf 1603}, 176 (2016)
  [arXiv:1512.05327 [hep-ph]].
  
\bibitem{Gupta:2015zzs} 
  R.~S.~Gupta, S.~Jager, Y.~Kats, G.~Perez and E.~Stamou,
  arXiv:1512.05332 [hep-ph].
  
\bibitem{Kobakhidze:2015ldh}
  A.~Kobakhidze, F.~Wang, L.~Wu, J.~M.~Yang and M.~Zhang,
  Phys.\ Lett.\ B {\bf 757} (2016) 92
  [arXiv:1512.05585 [hep-ph]].
  
\bibitem{Falkowski:2015swt} 
  A.~Falkowski, O.~Slone and T.~Volansky,
  JHEP {\bf 1602}, 152 (2016)
  [arXiv:1512.05777 [hep-ph]].
  
\bibitem{Benbrik:2015fyz} 
  R.~Benbrik, C.~H.~Chen and T.~Nomura,
  Phys.\ Rev.\ D {\bf 93}, no. 5, 055034 (2016)
  [arXiv:1512.06028 [hep-ph]].
  
\bibitem{Wang:2015kuj}
  F.~Wang, L.~Wu, J.~M.~Yang and M.~Zhang,
  Phys.\ Lett.\ B {\bf 759} (2016) 191
  [arXiv:1512.06715 [hep-ph]].
  
  
  
\bibitem{Dev:2015isx} 
  P.~S.~B.~Dev and D.~Teresi,
  arXiv:1512.07243 [hep-ph].

\bibitem{Allanach:2015ixl} 
  B.~C.~Allanach, P.~S.~B.~Dev, S.~A.~Renner and K.~Sakurai,
  Phys.\ Rev.\ D {\bf 93}, no. 11, 115022 (2016)
  [arXiv:1512.07645 [hep-ph]].
  
\bibitem{Cheung:2015cug} 
  K.~Cheung, P.~Ko, J.~S.~Lee, J.~Park and P.~Y.~Tseng,
  arXiv:1512.07853 [hep-ph].

  
\bibitem{Wang:2015omi}
  F.~Wang, W.~Wang, L.~Wu, J.~M.~Yang and M.~Zhang,
  arXiv:1512.08434 [hep-ph].
  
\bibitem{Chiang:2015tqz} 
  C.~W.~Chiang, M.~Ibe and T.~T.~Yanagida,
  JHEP {\bf 1605}, 084 (2016)
  [arXiv:1512.08895 [hep-ph]].
  
\bibitem{Huang:2015svl} 
  X.~J.~Huang, W.~H.~Zhang and Y.~F.~Zhou,
  Phys.\ Rev.\ D {\bf 93}, 115006 (2016)
  [arXiv:1512.08992 [hep-ph]].
  
\bibitem{Kanemura:2015bli} 
  S.~Kanemura, K.~Nishiwaki, H.~Okada, Y.~Orikasa, S.~C.~Park and R.~Watanabe,
  arXiv:1512.09048 [hep-ph].
  
\bibitem{Nomura:2016fzs} 
  T.~Nomura and H.~Okada,
  Phys.\ Lett.\ B {\bf 755}, 306 (2016)
  [arXiv:1601.00386 [hep-ph]].
  
\bibitem{Ko:2016lai} 
  P.~Ko, Y.~Omura and C.~Yu,
  JHEP {\bf 1604}, 098 (2016)
  [arXiv:1601.00586 [hep-ph]].

  
\bibitem{Ko:2016wce} 
  P.~Ko and T.~Nomura,
  Phys.\ Lett.\ B {\bf 758}, 205 (2016)
  [arXiv:1601.02490 [hep-ph]].
  
\bibitem{Dorsner:2016ypw} 
  I.~Dorsner, S.~Fajfer and N.~Kosnik,
  Phys.\ Rev.\ D {\bf 94}, no. 1, 015009 (2016)
  [arXiv:1601.03267 [hep-ph]].
  
\bibitem{Nomura:2016seu} 
  T.~Nomura and H.~Okada,
  arXiv:1601.04516 [hep-ph].
  
\bibitem{Han:2016bvl} 
  X.~F.~Han, L.~Wang and J.~M.~Yang,
  Phys.\ Lett.\ B {\bf 757}, 537 (2016)
  [arXiv:1601.04954 [hep-ph]].
  
\bibitem{Belanger:2016ywb} 
  G.~Belanger and C.~Delaunay,
  arXiv:1603.03333 [hep-ph].
 
 
  
 
 
\bibitem{ATLAS:2016eeo} 
  The ATLAS collaboration [ATLAS Collaboration],
  ATLAS-CONF-2016-059.
 
\bibitem{CMS:2016crm} 
  CMS Collaboration [CMS Collaboration],
  CMS-PAS-EXO-16-027.
 
 
 
\bibitem{CMS:2014xfa} 
  V.~Khachatryan {\it et al.} [CMS and LHCb Collaborations],
  Nature {\bf 522}, 68 (2015)
  [arXiv:1411.4413 [hep-ex]].

\bibitem{Bobeth:2013uxa} 
  C.~Bobeth, M.~Gorbahn, T.~Hermann, M.~Misiak, E.~Stamou and M.~Steinhauser,
  Phys.\ Rev.\ Lett.\  {\bf 112}, 101801 (2014)
  [arXiv:1311.0903 [hep-ph]].


 
\bibitem{Bauer:2015boy} 
  M.~Bauer and M.~Neubert,
  Phys.\ Rev.\ D {\bf 93}, no. 11, 115030 (2016)
  [arXiv:1512.06828 [hep-ph]].
 
\bibitem{Murphy:2015kag} 
  C.~W.~Murphy,
  Phys.\ Lett.\ B {\bf 757}, 192 (2016)
  [arXiv:1512.06976 [hep-ph]].
 
\bibitem{Chao:2015nac} 
  W.~Chao,
  arXiv:1512.08484 [hep-ph].
 
\bibitem{Hati:2016thk} 
  C.~Hati,
  Phys.\ Rev.\ D {\bf 93}, no. 7, 075002 (2016)
  [arXiv:1601.02457 [hep-ph]].
 
\bibitem{Deppisch:2016qqd} 
  F.~F.~Deppisch, S.~Kulkarni, H.~Pas and E.~Schumacher,
  Phys.\ Rev.\ D {\bf 94}, no. 1, 013003 (2016)
  [arXiv:1603.07672 [hep-ph]].
  
\bibitem{Dey:2016sht} 
  U.~K.~Dey, S.~Mohanty and G.~Tomar,
  arXiv:1606.07903 [hep-ph].

\bibitem{DiIura:2016wbx} 
  A.~Di Iura, J.~Herrero-Garcia and D.~Meloni,
  arXiv:1606.08785 [hep-ph].




\bibitem{Aad:2015gba}
  G.~Aad {\it et al.} [ATLAS Collaboration],
  Eur.\ Phys.\ J.\ C {\bf 76} (2016) no.1,  6
  [arXiv:1507.04548 [hep-ex]].
  
 
\bibitem{CMS:2014ega} 
  CMS Collaboration [CMS Collaboration],
  CMS-PAS-HIG-14-009.
  
  
\bibitem{Adam:2013mnn} 
  J.~Adam {\it et al.} [MEG Collaboration],
  Phys.\ Rev.\ Lett.\  {\bf 110}, 201801 (2013)
  [arXiv:1303.0754 [hep-ex]].
 
\bibitem{Baek:2015mea} 
  S.~Baek and K.~Nishiwaki,
  Phys.\ Rev.\ D {\bf 93}, no. 1, 015002 (2016)
  [arXiv:1509.07410 [hep-ph]].
  
\bibitem{Baek:2016kud} 
  S.~Baek, T.~Nomura and H.~Okada,
  arXiv:1604.03738 [hep-ph].
 
   \bibitem{Chen:2002bq} 
  C.~H.~Chen and C.~Q.~Geng,
  Nucl.\ Phys.\ B {\bf 636}, 338 (2002)
  [hep-ph/0203003].
  
\bibitem{Chen:2002zk} 
  C.~H.~Chen and C.~Q.~Geng,
  Phys.\ Rev.\ D {\bf 66}, 094018 (2002)
  [hep-ph/0209352].
  
 \bibitem{Altmannshofer:2008dz} 
  W.~Altmannshofer, P.~Ball, A.~Bharucha, A.~J.~Buras, D.~M.~Straub and M.~Wick,
  JHEP {\bf 0901}, 019 (2009)
  [arXiv:0811.1214 [hep-ph]].
  
 \bibitem{Egede:2010zc} 
  U.~Egede, T.~Hurth, J.~Matias, M.~Ramon and W.~Reece,
  JHEP {\bf 1010}, 056 (2010)
  [arXiv:1005.0571 [hep-ph]].
 
  
\bibitem{Hiller:2003js} 
  G.~Hiller and F.~Kruger,
  Phys.\ Rev.\ D {\bf 69}, 074020 (2004)
  [hep-ph/0310219].

 
 \bibitem{Khachatryan:2015kon} 
  V.~Khachatryan {\it et al.} [CMS Collaboration],
  Phys.\ Lett.\ B {\bf 749}, 337 (2015)
  [arXiv:1502.07400 [hep-ex]].
  
\bibitem{Aad:2015gha} 
  G.~Aad {\it et al.} [ATLAS Collaboration],
  JHEP {\bf 1511}, 211 (2015)
  [arXiv:1508.03372 [hep-ex]].

  

  

  
   

  
  
  
  
  


  
  
  

\end{thebibliography}
\end{document}